\documentclass[sigconf,screen]{acmart}
\usepackage{graphicx}
\usepackage{subcaption}
\usepackage{stfloats}
\usepackage{csquotes}
\usepackage{balance}

\setcopyright{acmcopyright}
\copyrightyear{2026}
\acmYear{2026}
\acmDOI{XXXXXXX.XXXXXXX}
\acmConference[WiPSCE '26]{The 20th WiPSCE Conference on Primary and Secondary Computing Education}{March 11 -- March 13, 2026}{Aachen, GER}
\acmPrice{XX.XX}
\acmISBN{XXX-X-XXXX-XXXX-X/XX/XX}

\begin{document}

\title{Strategy-Oriented Feedback for Fostering Systematic Problem-Solving in Machine Learning Education}

\author{Clemens Witt}
\orcid{0009-0005-8160-4029}
\email{clemens.witt@tu-dresden.de}
\author{Mareen Grillenberger}
\orcid{0000-0002-8477-1464}
\email{mareen.grillenberger@tu-dresden.de}
\affiliation{%
  \institution{TUD Dresden University of Technology}
  \streetaddress{Nöthnitzer Str. 46}
  \city{Dresden}
  \country{Germany}
  \postcode{01187}
}

\author{Thiemo Leonhardt}
\email{leonhardt@cs.rwth-aachen.de}
\orcid{0000-0003-4725-9776}
\affiliation{
  \institution{RWTH Aachen University}
  \streetaddress{Mies-van der Rohe Str. 15}
  \city{Aachen}
  \country{Germany}
  \postcode{52074}
}

\author{Erik Marx}
\orcid{0000-0002-5918-804X}
\email{erik.marx@tu-dresden.de}
\additionalaffiliation{%
  \institution{TUD Dresden University of Technology}
  \streetaddress{Nöthnitzer Str. 46}
  \city{Dresden}
  \country{Germany}
  \postcode{01187}
}
\affiliation{%
  \institution{Center for Scalable Data Analytics and Artificial Intelligence (ScaDS.AI)}
  \city{Dresden/Leipzig}
  \country{Germany}
}

\renewcommand{\shortauthors}{Clemens Witt, Thiemo Leonhardt, Erik Marx, and Mareen Grillenberger}

\begin{abstract}
    Enabling students to develop systematic problem-solving strategies is a central goal in computing education and of particular relevance in the emerging field of machine learning (ML) education.
    While exploratory approaches are common in ML learning tasks, fostering the development and persistence of structured problem-solving strategies remains challenging, as these demand considerable metacognitive regulation and persistence, causing learners to often revert to exploratory trial-and-error behavior.
    To address this challenge, we augmented a digital puzzle-based learning game for decision tree construction with an adaptive feedback module generating individualized messages based on the continuous evaluation of learners’ problem-solving strategies. Building on an earlier baseline study, the present work investigates how this strategy-oriented feedback shapes students’ problem-solving processes. 
    For this purpose, screencast video data and gameplay logs ($N=205$, $\approx$55 hours of gameplay footage) are used to enable fine-grained insights into learners’ strategic behavior, its persistence, and transitions.
    The findings demonstrate how strategy-oriented feedback can support the development of structured problem-solving skills in decision tree construction and inform the design of ML learning environments that foster transferable competencies in secondary computing education.
\end{abstract}



\maketitle

\section{Introduction}
The development of problem-solving competencies is widely recognized as a central objective of computing education \cite{groverComputationalThinkingK122013}.
Within the rapidly expanding field of education in Artificial Intelligence (AI) and Machine Learning (ML), these competencies take on particular importance.
In contrast to traditional domains of computing education, learners are often not primarily confronted with well-defined problems and deterministic solution paths, but must work inductively with data, manage uncertainty, and critically reflect on model behavior \cite{longWhatAILiteracy2020,tedreCT202021a,touretzkyMachineLearningFive2022}.
As these practices often involve considerable cognitive and metacognitive demands, and learners frequently lack appropriate heuristics, they tend in practice to resort to exploratory tinkering strategies.
While such strategies may yield short-term progress, they rarely support the development of sustainable and transferable problem-solving strategies \cite{wittGuessworkGamePlan2024a}.
Prior research in computing education has repeatedly demonstrated that, although structured problem-solving strategies are widely acknowledged as a key educational goal, they are rarely fostered systematically in classroom practice \cite{groverComputationalThinkingK122013}.
For learning contexts in ML education, this underscores a fundamental pedagogical challenge: enabling learners to develop strategies for addressing ML-related tasks and encouraging them to sustain these strategies consistently throughout the learning process.
Digital game-based learning environments hold particular promise in this regard, as they allow the real-time capture of learners’ problem-solving processes and the targeted integration of adaptive support mechanisms without interrupting the flow of learning \cite{dicerboFutureAssessmentTechnologyRich2016}.
However, research explicitly targeting the levels of process and self-regulation, e.g., feedback addressing strategic patterns of action and their deliberate regulation, remains limited \cite{hattiePowerFeedback2007a,narcissFeedbackStrategiesInteractive2008a}.
Especially in the context of AI and ML education, systematic investigations and established design concepts for such feedback are still lacking \cite{longWhatAILiteracy2020,ngArtificialIntelligenceAI2024}.
This study provides an exploratory contribution by addressing this gap through the development and empirical investigation of a strategy-oriented feedback system for the digital learning game \textit{Match the Monkeys}, situated in the context of decision tree learning.
The system is based on a multimodal ML model specifically designed to detect students’ problem-solving strategies in real time, thereby enabling the adaptive provision of individualized feedback across the entire gameplay process.
This approach creates new possibilities for fostering structured problem-solving strategies in a targeted and sustained manner.
The investigation is guided by the following research question:

\begin{quote}
    \textit{What influence does adaptive, strategy-oriented feedback have on the development and persistence of structured problem-solving strategies in secondary school students when constructing decision trees in a self-directed learning setting?}    
\end{quote}

With this investigation, the study contributes to empirically substantiating the potential of strategy-oriented feedback in ML education and offers new perspectives for the design of digital learning environments that support the acquisition of sustainable and transferable problem-solving competencies.
The subsequent sections first outline the theoretical background and relevant prior work, followed by a description and discussion of the developed feedback system, methodological approach, and empirical findings.
Building on this, the pedagogical implications for the design of future learning environments and the advancement of feedback systems in ML education are elaborated.
\section{Theoretical Background}
As outlined above, ML problem-solving differs fundamentally from the problem-solving approaches typically emphasized in traditional computing education.
Learners therefore enter ML learning tasks with fewer established problem-solving routines and a greater reliance on developing and testing assumptions, shaping the conditions of their problem-solving processes.
To investigate these processes empirically, digital learning environments provide opportunities to capture them as they unfold, with \textit{stealth assessment} methods linking theoretical models of problem-solving to empirical evidence from interaction data.
Strategy-oriented feedback constitutes a complementary perspective by actively shaping how problem-solving strategies are developed and regulated during learning.
Together, these three perspectives form the theoretical foundation of this study.

\subsection{Problem-Solving in ML Education}
Problem-solving is a central aspect in both computer science as a didactic discipline and in professional practice.
It often involves the integration of diverse domain-specific concepts and skills, making its structured development a longstanding goal in computing education \cite{wooProblemSolvedHow2022b}.
While domain knowledge is typically taught in a systematic fashion, problem-solving skills are often acquired implicitly and without formal scaffolding \cite{hazzanProblemSolvingStrategies2020}.

Published in 1945, Pólya's four-step heuristic \textit{(understanding the problem, devising a plan, executing the plan, and reflecting on the solution)} provided the first systematic framework for fostering problem-solving competence in educational contexts \cite{polyaHowSolveIt1945}.
Though originally developed in mathematics, this heuristic influenced early CS education by offering parallels to algorithm design and implementation \cite{wooProblemSolvedHow2022b}. 
However, in CS, Pólya's original model was soon expanded, as problem-solving is often more iterative, exploratory, and incremental, necessitating the inclusion of practices such as abstraction, modeling, algorithmic thinking, and metacognitive reflection \cite{maharaniPROBLEMSOLVINGCONTEXT2019}.
From the 1980s onward, constructionist perspectives, especially Papert's conception of the computer as a \textit{thinking tool}, framed programming as a medium for engaging learners in creative, and reflective problem-solving. 
Learning environments such as LOGO were designed to enable learners to experience heuristic problem-solving processes in an exploratory, hands-on manner 
\cite{papertConstructionismNewOpportunity1986}.
In the 2000s, the paradigm of Computational Thinking (CT) expanded the educational focus beyond programming, introducing computing-related problem-solving as a fundamental skill for all learners.
In her seminal essay, \citet{wingComputationalThinking2006a} framed CT as a universally applicable way of thinking, arguing that its principles should be taught alongside reading, writing, and arithmetic.
In a subsequent publication \cite{wingComputationalThinkingThinking2008}, she refined this perspective by identifying abstraction as the defining characteristic of CT, emphasizing its central role in managing complexity across problem domains.
Building on Wing's foundational work, later contributions (e.g. \cite{brennanNewFrameworksStudying2012,shuteDemystifyingComputationalThinking2017}) expanded the model by operationalizing CT into a set of core practices and cognitive skills.
These include \textit{decomposition}, \textit{algorithm design}, \textit{iteration}, \textit{debugging}, and \textit{generalization}, which serve to concretize CT for educational practice and assessment.

The rise of machine learning (ML) in computing practice has introduced a qualitatively different model of problem-solving.
Traditional CT follows a predominantly deductive, top-down logic: human-formulated problems are abstracted, decomposed, and translated into algorithms \cite{nowackModelbasedThinkingPractice2014}. Inductive elements (e.g., iteration and debugging) emerge later in the process \cite{brennanNewFrameworksStudying2012,shuteDemystifyingComputationalThinking2017}.
In contrast, ML problem-solving often begins inductively, with problems emerging from data. Learners must identify suitable target variables, data representations, and labels before even formulating the problem fully \cite{passiProblemFormulationFairness2019}.
Deductive strategies typically emerge only at later stages, for instance during model evaluation or when aligning model metrics with external goals or constraints \cite{liuReimaginingMachineLearning2023}.
Widely-used educational ML process models such as the ML workflow model \cite{zimmermanTeachingAIExploring2018} start with data collection and curation, underscoring the bottom-up nature of problem formulation in ML.
These structural differences prompted a reevaluation of CT in light of ML.
Several publications have emphasized the need to broaden CT beyond algorithm design, highlighting the importance of practices for handling complex datasets, probabilistic reasoning to address uncertainty, and ethical reflection in the design and deployment of models \cite{groverComputationalThinkingCompetency2018,shuteDemystifyingComputationalThinking2017,weintropDefiningComputationalThinking2016}.
A comprehensive reframing of CT in the context of ML is presented in \textit{Computational Thinking 2.0} (CT 2.0) by \citet{tedreCT202021a}.
Like Wing's original vision for CT, CT 2.0 offers a didactic orientation framework that explicitly differentiates ML-specific practices from classic CT elements:

\begin{itemize}
    \item \textit{Data work} (collection, curation, labeling) replaces problem definition as the starting point,
    \item \textit{Modeling through training and tuning} substitutes for algorithmic design,
    \item \textit{Evaluation} involves heuristic metrics rather than formal correctness proofs,
    \item \textit{Optimization and debugging} are guided by performance feedback,
    \item and \textit{Critical reflection} addresses bias and opacity in learned models.
\end{itemize}

CT 2.0 provides a systematic elaboration that conceptualizes these practices in relation to classical CT dimensions (e.g., abstraction, generalization, and debugging), thereby facilitating curricular alignment and pedagogical accessibility.
Since its introduction, the framework has been established as a central reference point in AI and ML education, informing the design of competency models, teaching strategies, and empirical studies (e.g. \cite{conwayHCAIBlockModel2024,morales-navarroConstructionistApproachesLearning2023}).
Despite its contributions, CT 2.0 also reveals notable limitations.
These include insufficient attention to data lifecycle practices such as quality, provenance, and fairness \cite{olariDatarelatedPracticesCreating2024a}, limited support strategies for probabilistic and statistical reasoning \cite{fleischerTEACHINGLEARNINGCONSTRUCT2024}, and only limited consideration of mechanisms for transparency and auditing in AI/ML education \cite{morales-navarroUnpackingApproachesLearning2024}.
Further questions remain regarding pathways for teacher professionalization and the development of valid assessment instruments for ML-specific competencies \cite{kimSystematicReviewEvaluation2025}.

\subsection{Stealth Assessment in Digital Learning Environments}
A key feature of digital, game-based learning environments is their potential to design learning experiences that are both effective and motivating \cite{emersonMultimodalLearningAnalytics2020a}.
For this reason, they provide a particularly suitable environment for formative assessment, as learning can be captured and evaluated in real time without interrupting the learner's immersion or the natural flow of gameplay \cite{dicerboFutureAssessmentTechnologyRich2016}.
\textit{Stealth assessment} in this context refers to a formative approach in which the collection and evaluation of learning outcomes is seamlessly and unobtrusively embedded into the digital environment \cite{shuteStealthAssessmentMeasuring2013}.
The concept was originally proposed by \citet{shuteSimplyAssessment2009}, based on the framework of Evidence-Centered Design (ECD) \cite{mislevyFocusArticleStructure2003}.
Within this framework, the \textit{competency model} defines the target knowledge and skills, the \textit{evidence model} specifies the observable actions from which competencies can be inferred, and the \textit{task model} describes activities that elicit these actions \cite{mislevyDesignDiscoveryEducational2012}.

While evidence models in the sense of ECD have traditionally been developed manually by experts, there is a growing line of research that explores the use of ML techniques to construct such models automatically.
This shift is driven both by the fine-grained interaction data available from gameplay logs in digital learning games and by the substantial feature-engineering effort required for manual construction of evidence models \cite{guptaMultimodalMultiTaskStealth2021,minDeepStealthGameBasedLearning2020}.
To date, much of this research has focused on predicting learning outcomes and performance data from gameplay interactions, with the goal of enabling individualized support and adaptive feedback (e.g. \cite{guptaMultimodalMultiTaskStealth2021, hendersonEnhancingStealthAssessment2022}).
Beyond outcome prediction, more recent approaches emphasize the analysis of learning processes and problem-solving strategies, for example by modeling how learners explore, revise, and justify their solutions \cite{akramImprovingStealthAssessment,guptaMultimodalMultiTaskStealth2021}.
More recently, multimodal assessment approaches have emerged that integrate additional data sources such as written reflections, physiological signals, or socio-emotional behaviors.
By combining heterogeneous data, these approaches allow for a more differentiated modeling of cognitive, emotional, and metacognitive processes \cite{guptaMultimodalMultiTaskStealth2021,hendersonEnhancingStealthAssessment2022}.

\subsection{Strategy-Oriented Feedback Design}
Feedback is widely acknowledged as a decisive factor in effective learning, as it enables learners to recognize discrepancies between their current level of understanding and defined learning objectives, and to systematically work toward overcoming them \cite{hattiePowerFeedback2007a}.
Within digital learning environments, feedback fulfills a dual function: it provides orientation for learning processes while simultaneously fostering reflection and offering actionable guidance \cite{maierPersonalizedFeedbackDigital2022a}.
In particular, within complex problem-solving contexts, feedback assumes a formative role by rendering strategies visible and modifiable rather than merely evaluating outcomes \cite{shuteFocusFormativeFeedback2008a}.
Empirical studies indicate that feedback is most effective when it is specific, timely, and closely aligned with learning goals, whereas generic or delayed responses exhibit markedly reduced efficacy \cite{morrisonDesigningEffectiveInstruction2019}.

Feedback is especially beneficial when it delivers constructive indications for improving problem-solving strategies instead of exclusively identifying errors.
Three interrelated dimensions are central in this regard: \textit{cognitive feedback}, which targets disciplinary content; \textit{metacognitive feedback}, which promotes reflection on the learner's approach; and \textit{motivational feedback}, which strengthens persistence and self-efficacy \cite{narcissFeedbackStrategiesInteractive2008a,shuteFocusFormativeFeedback2008a}.
Furthermore, effective feedback must be adapted to individual learner characteristics, considering not only their domain-specific proficiency but also their emotional and motivational states \cite{linnenbrink-garciaAdaptiveMotivationEmotion2016}.
Beyond immediate task performance, formative feedback contributes to the systematic transformation of unsystematic, exploratory approaches into reflective, structured, and sustainable problem-solving strategies \cite{wooDigitalGameBasedLearning2014}.

In current research, automatic feedback generation through Large Language Models (LLMs) is attracting increasing attention.
Contemporary LLMs are capable of generating differentiated and context-sensitive feedback that addresses both cognitive and metacognitive processes \cite{steinertHarnessingLargeLanguage2024}.
However, a key challenge lies in their inconsistent grounding in observable learner artifacts: responses may contain claims not substantiated by the data or may contradict it \cite{qinjinjiaAssessingFaithfulnessLLMgenerated2024}.
Accordingly, the development of LLM-based feedback systems requires strict artifact-referenced design and systematic validation prior to educational deployment.

Digital game-based environments offer additional potential, as feedback can be embedded directly into the activity flow.
In this context, immediate responses have been shown to increase motivation and facilitate iterative strategy refinement \cite{wooDigitalGameBasedLearning2014}, while excessive information can inhibit autonomous learning \cite{kalyugaEvaluatingManagingCognitive2009}.
Effective feedback systems therefore must balance support and autonomy, for example through adaptive pacing that dynamically calibrates task difficulty and assistance to learner competence, thereby promoting flow and preventing under- or over-challenge \cite{koskinenStrengthDirectionDifficulty2023}.
This balance can be achieved through concise, context-specific micro-hints delivered just-in-time and elaborated only when difficulties persist \cite{kalyugaEvaluatingManagingCognitive2009}.
Thus, effective strategy-oriented feedback design can be characterized as the integration of preventive and corrective elements that foster self-regulation by directing attention to learners' problem-solving strategies and encouraging their systematic refinement \cite{butlerFeedbackSelfRegulatedLearning1995}.
\section{Preliminary Work}
\begin{figure}[h]
    \centering
    \includegraphics[width=\linewidth]{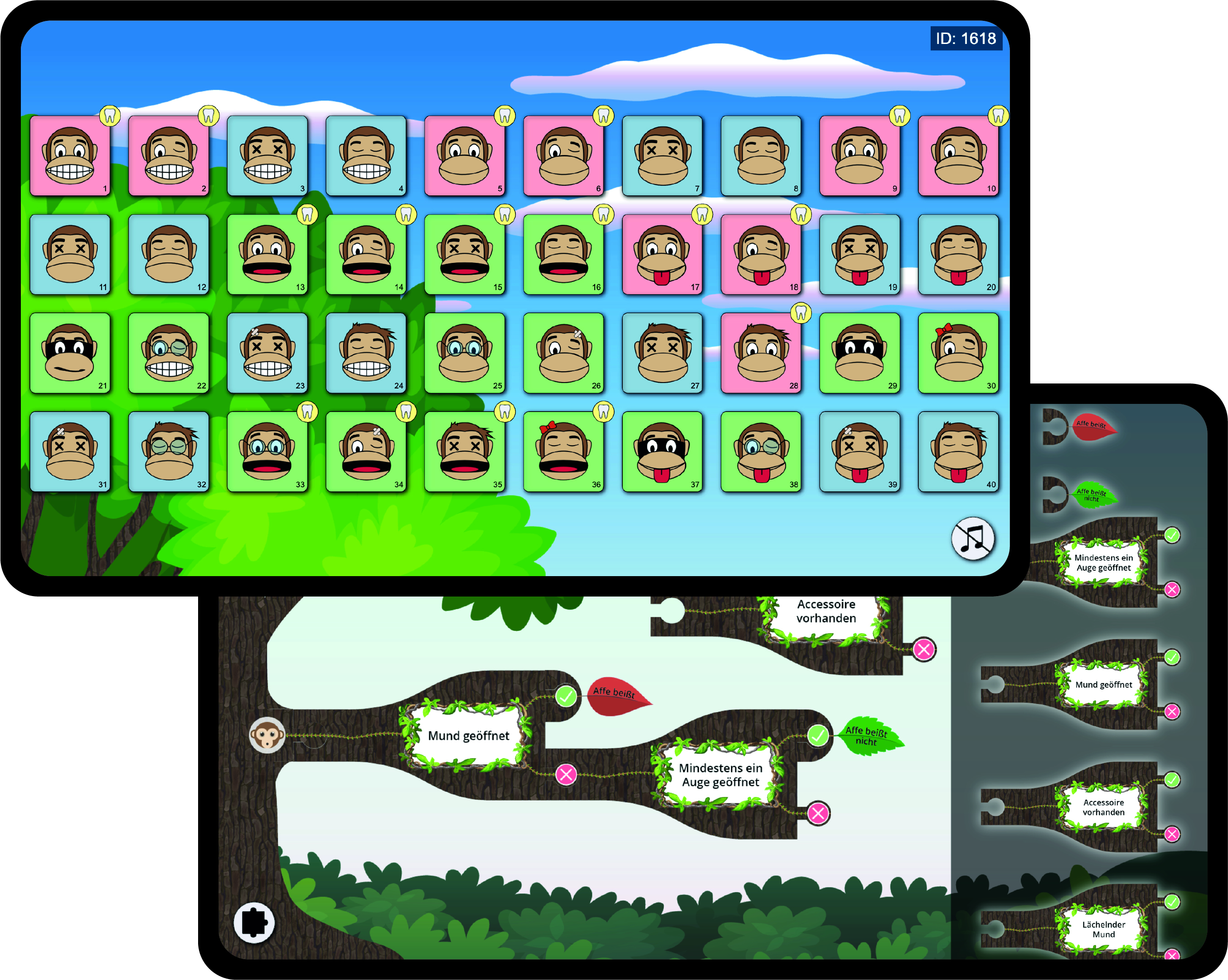}
    \caption{Screenshot of the \textit{Match the Monkeys learning} game interface showing the puzzle-based decision tree construction area (bottom) and the color-coded feedback display (top)}
    \label{fig:userinterface}
\end{figure}

This study builds on previous work on secondary school students' problem-solving during the self-regulated construction of decision trees in the digital learning game \textit{Match the Monkeys} (see Fig. \ref{fig:userinterface}). In this line of research, we developed and validated a coding scheme to capture learners' problem-solving strategies (see Fig. \ref{fig:coding-guide}), conducted an empirical study to analyze their distribution and persistence, and introduced a ML model for automatically classifying these strategies. The following sections summarize this preliminary work, which provides both the empirical basis and the methodological foundation for the present study.

\subsection{Learning Game \textit{Match the Monkeys}}

To investigate students' problem-solving strategies when constructing decision trees, we developed the digital learning game \textit{Match the Monkeys} \cite{wittGuessworkGamePlan2024a}. 
The game is based on the \textit{Good Monkey, Bad Monkey Game} learning activity originally developed as part of the \textit{AI-Unplugged} framework by \citet{lindnerUnpluggedActivitiesContext2019} – a collection of educational resources designed to provide accessible entry points into the field of AI. 
The objective of the game is to correctly distinguish between biting and non-biting monkeys based on their visual features by constructing a decision tree.
To this end, learners select from a set of filter elements in the form of puzzle pieces and assemble them into a functional classification model.
With the help of color-coded corrective feedback displayed on a dedicated screen, they iteratively adapt their model to the given data, gradually improving the classification accuracy of their decision tree.

\subsection{Category System Development}

\begin{figure}[!h]
    \centering
    \includegraphics[width=\linewidth]{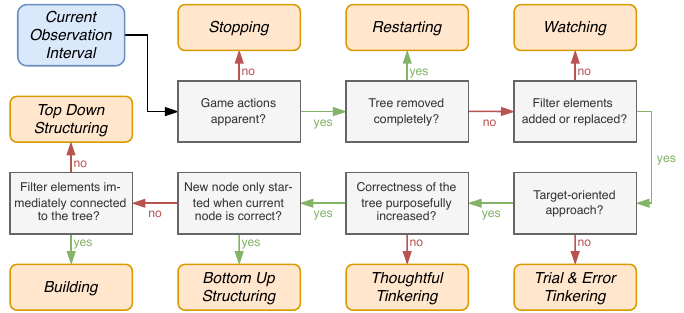}
     \caption{Coding guide for identifying student problem-solving strategies during gameplay, developed and validated in the baseline study $(\alpha = 0.75)$}
    \label{fig:coding-guide}
\end{figure}

To analyze students' problem-solving behaviors in the \textit{Match the Monkeys} learning game, we developed a coding scheme derived from a literature review of established problem-solving strategies in programming education \cite{wittGuessworkGamePlan2024a}.
The resulting categories capture typical behavior patterns observed in programming-related learning tasks, such as exploratory and targeted tinkering (\textit{Thoughtful Tinkering}, \textit{Trial \& Error Tinkering}, and \textit{Watching}), as well as structured problem-solving approaches (\textit{Bottom-Up Structuring}, \textit{Top-Down Structuring}, and \textit{Building}).
These strategies were contextually adapted to decision tree construction, a domain that – similar to programming – requires iterative decomposition of complex tasks and reflective adaptation based on feedback.
The interactive and visual game environment of the \textit{Match the Monkeys} learning game provided a controlled setting to observe and analyze these behaviors without requiring prior formal knowledge of decision tree learning.
The coding scheme was validated in a coding training with five coders applying quantitative content analysis \cite{fruhInhaltsanalyse2017} in 30-second observation intervals, achieving an intercoder reliability of Krippendorff's $\alpha = 0.75$.

\subsection{Insights into Students' Problem-Solving} \label{pre-study}
Based on the validated coding scheme, we analyzed secondary school students' gameplay in the \textit{Match the Monkeys} learning game \cite{wittGuessworkGamePlan2024a}.
The results show a clear predominance of exploratory problem-solving strategies: \textit{Trial \& Error Tinkering} and \textit{Thoughtful Tinkering} together accounted for the vast majority of observation intervals, indicating that most learners progressed through iterative experimentation and incremental adjustments.
Structured approaches (\textit{Top-Down Structuring}, \textit{Bottom-Up Structuring}, \textit{Building}) appeared far less frequently. Two key challenges emerged: only few students were able to independently develop structured strategies, and even when initiated, such strategies were seldom sustained across consecutive intervals, with learners often reverting to tinkering.
These findings underline that while students intuitively rely on exploratory tinkering, generating and maintaining structured approaches remains a considerable challenge. This insight not only validates the relevance of distinguishing between different problem-solving categories but also highlights a clear pedagogical need: supporting learners in stabilizing structured strategies. Addressing this challenge directly motivated the design of the feedback system investigated in the present study.

\subsection{Multimodal Strategy Classification Model} \label{strategy-classification-model}
To enable the automatic assessment of students' problem-solving strategies in the \textit{Match the Monkeys} learning game, we developed a ML-based strategy recognition model \cite{wittMultimodalLateFusion2026} that builds directly on the validated coding scheme described above. The model was designed to go beyond conventional log-based approaches by integrating complementary data sources: gameplay screencasts and symbolic representations of learners' decision trees. This multimodal integration was necessary because screencasts provide access to subtle behavioral cues such as hesitation, pacing, or the sequence of puzzle manipulations, while symbolic action sequences capture the structural quality of the constructed decision tree models. Both modalities together allow a more valid operationalization of the strategy categories than either source alone.
For model training, the original coding scheme was adapted to the constraints of automatic classification. While the full scheme distinguished between six problem-solving strategies, only three target classes were used: \textit{Thoughtful Tinkering}, \textit{Trial \& Error Tinkering}, and \textit{Structured Problem-Solving}. The two tinkering categories were retained separately due to their prevalence and theoretical relevance, whereas the structured strategies (\textit{Top-Down Structuring}, \textit{Bottom-Up Structuring}, and \textit{Building}) were combined. This consolidation was necessary due to the limited representation of these structured approaches in the dataset. Although they differ in some respects, their observable interaction patterns in the game do not diverge enough to justify separate model classes, given the risk of class imbalance and unreliable model training.
The model was trained on 149 annotated gameplay sessions ($\approx$30 hours of gameplay footage) and achieved an overall F1-score of 0.754, with particularly strong performance in recognizing structured problem-solving strategies (F1=0.88). 
These results demonstrate that the multimodal model architecture can reliably distinguish between exploratory and structured problem-solving behavior during decision tree construction in the \textit{Match the Monkeys} learning game, thereby providing a robust foundation for future applications in adaptive, game-based assessment systems.
\section{Multimodal Feedback System Design}

\begin{figure}[h]
    \centering
    \includegraphics[width=\linewidth]{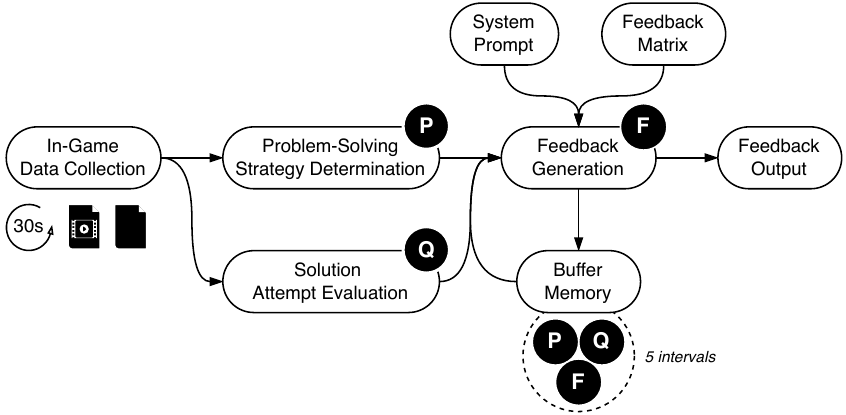}
    \caption{Real-time adaptive feedback pipeline based on strategy classification and solution accuracy evaluation}
    \label{fig:feedback-pipeline}
\end{figure}

The feedback system integrated into the \textit{Match the Monkeys} learning game leverages a large language model (LLM) to generate adaptive responses from multimodal data on learners’ problem-solving strategies and decision tree accuracy. 
A pilot with 10th-grade students ($N=19$) using the same data collection procedure as in the main study (see Section \ref{data-acquisition}) ensured clear feedback, alignment with observed strategies, and technical robustness. 
The following paragraphs outline the system’s inner workings and key design decisions. 
To promote transparency and reproducibility, the complete system prompt, the finalized feedback matrix, and the full source code are available via a dedicated OSF repository \cite{OSFRepository}.

\subsection{Feedback Generation Pipeline Overview}
\begin{figure}[h]
    \centering
    \includegraphics[width=.9\linewidth]{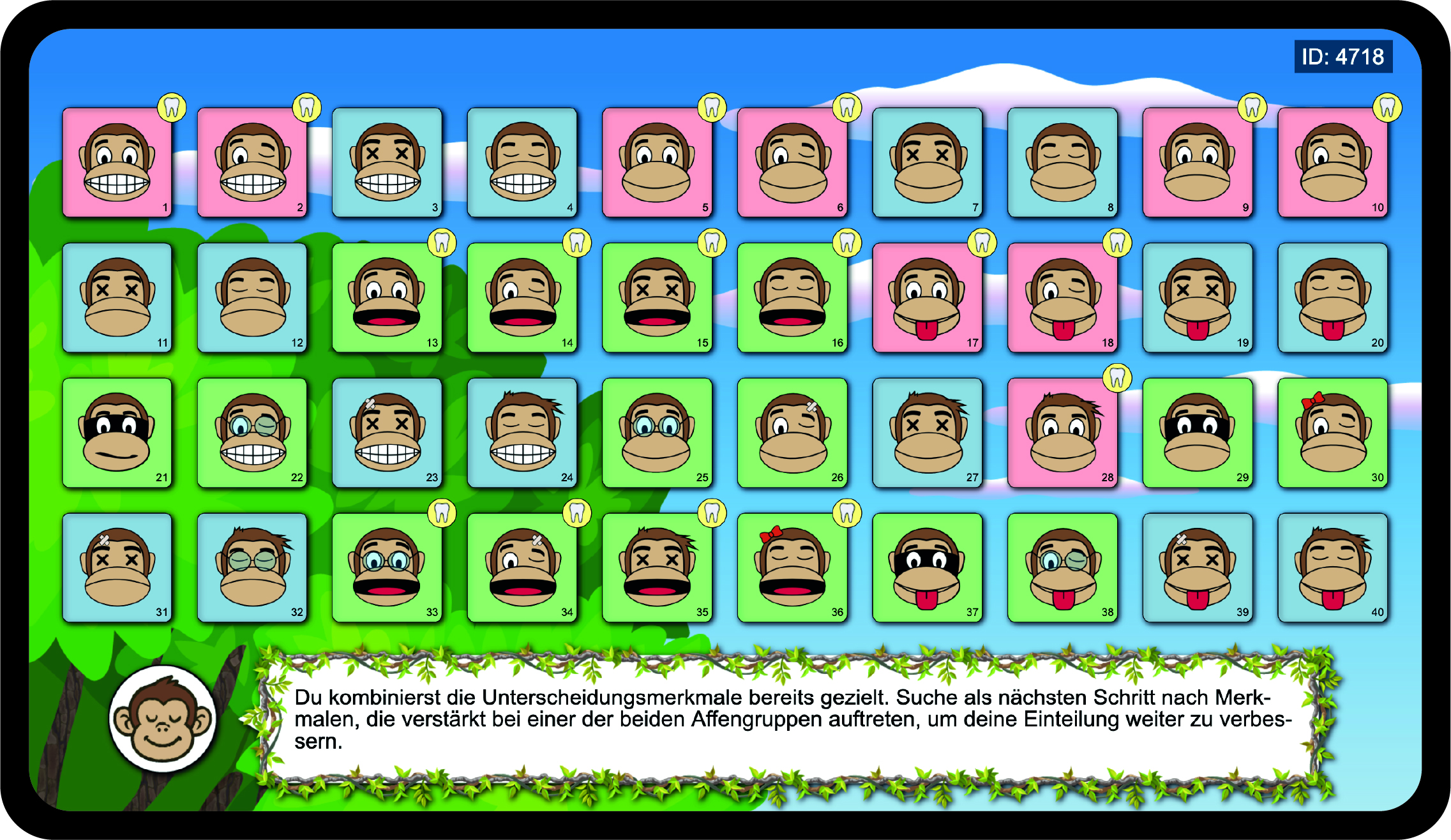}
    \caption{Feedback agent integrated into the \textit{Match the Monkeys} learning game, providing adaptive, strategy-oriented guidance during decision tree construction}
    \label{fig:feedback-design}
\end{figure}

In the game, adaptive feedback is generated in a multi-stage process, that draws on continuous data collection at 30-second observation intervals, incorporating both screencasts of the game interface and JSON-based logs of all student modifications to the evolving decision tree (see Fig. \ref{fig:feedback-pipeline}).
After the completion of each interval, two central dimensions of the student's problem-solving behavior are determined:
\textit{(1) The Problem-Solving Strategy}, classified by the multimodal ML model described in Section \ref{strategy-classification-model}, and \textit{(2) The Solution Attempt Accuracy}, operationalized as the current overall accuracy of the decision tree, computed within the game and subsequently categorized.
The combination of both dimensions is aligned with a specially designed feedback matrix, which specifies an appropriate feedback strategy for each observed manifestation.
Based on this, the LLM GPT-4.1 generates the concrete feedback message via an API request.
This model was selected as, at the time of development (early 2025), it provided advanced support for extended context windows \cite{openai2025gpt41} – an essential prerequisite for ensuring consistent feedback throughout longer gameplay sessions.
To ensure alignment with the predetermined feedback strategy, a specifically formulated system prompt guides the generation of the messages.
To further maintain coherence, the most recently determined strategies, accuracy values, and feedback messages are buffered in a sliding window across five observation intervals and incorporated as additional context in the generation process.
In this way, repetitions can be avoided and content consistency can be strengthened.
The final feedback messages are then delivered to the learners through the game's distributed display system (see. Fig. \ref{fig:feedback-design}).

\subsection{LLM System Prompt}
The system prompt constitutes the central control mechanism for feedback generation by the LLM.
It anchors the model in the context of the \textit{Match the Monkeys} learning game, defines the pedagogical objectives of feedback generation, and specifies the input parameters for each interval (\textit{problem-solving strategy}, \textit{decision tree accuracy}, \textit{prior feedback messages}, and the \textit{inferred feedback strategy}).
In addition, it establishes linguistic constraints, including concise formulations, a mandatory reference to observed in-game actions, and an objective and supportive tone without strategic or technical labeling.
The system prompt thus functions as a guiding structure that constrains the model's generative openness to ensure the generation of consistent, adaptive, and pedagogically aligned feedback messages.

\subsection{Feedback Matrix}
The feedback matrix constitutes a two-dimensional didactic reference structure for the generation of individualized feedback messages in the \textit{Match the Monkeys} learning game.
Along the strategy dimension, three categories are distinguished (\textit{Trial \& Error Tinkering}, \textit{Thoughtful Tinkering}, \textit{Structured Problem-Solving}).
Along the solution quality dimension, three levels of the overall accuracy of the evolving decision tree are differentiated: high ($\geq$80\%), medium ($\geq$50\%), and low ($<$50\%).
Each of the nine possible combinations is associated with specific guidelines that determine the content and pedagogical focus of the feedback to be generated.
The \textit{quality dimension} primarily varies the degree of specificity: high solution attempt accuracy results in confirmatory and stabilizing feedback, medium accuracy in guiding prompts to further develop the current approach, and low accuracy in more concrete next-step action suggestions.
In contrast, the \textit{strategy dimension} determines the pedagogical orientation of the feedback: when a \textit{Trial \& Error} strategy is detected, the goal is to reduce random attempts; with \textit{Thoughtful Tinkering}, the focus is on strengthening hypothesis formation and testing; and with \textit{Structured Problem-Solving}, the aim is to refine the learner's already deliberate approach.

\section{Methodology}
This section outlines the methodological framework of the present study. 
It describes the acquisition of gameplay data during introductory ML workshops, details the coder training and coding procedures employed to ensure reliability, and elaborates the statistical analyses conducted to examine both the distribution and the transitions of problem-solving strategies.

\subsection{Data Acquisition} \label{data-acquisition}
The recording of gameplay sessions with the extended feedback-enhanced version of the learning game \textit{Match the Monkeys} was conducted between June and August 2025 during a total of 11 introductory ML workshops in an extracurricular teaching laboratory.
In total, 230 students participated in the workshops, for whom this represented their first formal engagement with the topics of AI and ML.
The largest subgroup consisted of upper-secondary students ($N = 122$) in grades 8 ($N = 15$), 9 ($N = 81$), and 10 ($N = 26$).
In addition, one 7th-grade class from a lower-secondary comprehensive school ($N = 20$) and three 11th-grade classes from a technical-vocational secondary school ($N = 67$) participated.
Attendance generally took place as part of school-organized excursions within the subject of Computer Science.
An exception was a holiday program group of 21 mixed-grade academic track students who participated outside regular school activities.
In line with the preliminary study described in Section \ref{pre-study}, the learning game \textit{Match the Monkeys} was used as an entry point to the workshop module on models in the context of ML.
Learners played the game individually on two dedicated tablets provided by the laboratory.
Communication between participants was strictly prohibited by the workshop facilitators in order to capture the development of individual learning and gameplay strategies.
As in the preliminary study, students' gameplay sessions were documented through screencasts of the game interface.
The video data were stored under randomly generated session IDs and aggregated in a central database, resulting in a dataset comprising approximately 55 hours of recorded gameplay footage.

\subsection{Coder Training and Coding Process}
The analysis of problem-solving strategies in the recorded video material was carried out by means of quantitative content analysis \cite{fruhInhaltsanalyse2017}, drawing on the coding scheme developed in the preliminary study (see Fig. \ref{fig:coding-guide}).
Three coders participated in the analysis, each with prior training and application experience of the coding scheme from the preliminary study.
To further ensure consistency, an additional coder training was conducted prior to the initiation of the main coding phase.
For this purpose, five gameplay sessions were randomly selected from the pool of recorded sessions generated during the pilot of the adaptive feedback system.
Following the procedure of the preliminary study, these sessions were coded sequentially using the established coding scheme in observation intervals of 30 seconds.
Deviations were systematically reviewed under the supervision of the research lead, discussed in detail, and consensually revised where necessary.
Intercoder reliability, calculated using Krippendorff's Alpha, yielded an average value of $\alpha = 0.77$ across all training sessions, thereby slightly exceeding the level achieved in the preliminary study.
Subsequently, the recorded gameplay sessions were randomly allocated to the coders.
During individual coding, coders were instructed to refer ambiguities and doubtful cases to group discussion.
Such passages were reviewed collaboratively by the remaining coders and classified in a reasoned manner.
The final assignment of the respective problem-solving strategy was then determined through group consensus.

\subsection{Data Analysis}
The analysis of the coded gameplay sessions relied on a multi-step procedure integrating data preparation, non-parametric testing, and matrix-based modeling of strategy dynamics.
Prior to the statistical investigation, a data cleaning procedure was performed, in which game sessions were excluded that were suspected to contain technical malfunctions (e.g., connection interruptions between the game and the feedback tablet, incompletely recorded screencasts) or that exhibited atypical boundary conditions (e.g., tutorial duration <30 s, playtime <5 min or >25 min).

At the outset of the analysis, Shapiro–Wilk tests for normality were conducted for the relative shares of problem-solving strategies across the individual gameplay sessions.
As the assumption of normal distribution could not be confirmed for all strategies, group comparisons were subsequently carried out using the non-parametric Mann–Whitney U test.
To control for $\alpha$-error inflation in the case of multiple testing, the resulting $p$-values were adjusted using the Benjamini–Hochberg procedure \cite{benjaminiControllingFalseDiscovery1995}; results with corrected values of $q < 0.05$ were interpreted as statistically significant.
Correlations between relative strategy proportions and gameplay success were calculated using Spearman’s rank correlation coefficient.

For the analysis of strategy transition tendencies, the strategy sequences obtained during the coding process were used.
Based on the sequences of strategies in 30-second intervals, strategy transition matrices (STM) were constructed, in which rows represent the initial strategy and columns represent the subsequent strategy; the cell values indicate the relative frequencies of the observed transitions.
In this way, both persistence tendencies of individual strategies (main diagonal of the STM) as well as typical switching patterns between them can be systematically visualized.
Methodologically, the transitions per session were first captured in transition count matrices $M_s$ and then normalized row-wise, resulting in session-specific probability matrices $P_s$.
Each entry $P_s(i,j)$ thus describes the conditional probability of a transition from strategy $i$ to strategy $j$ within a session $s$.
The session-wise transition probabilities obtained in this manner were subsequently compared between the baseline and the feedback version using the Mann–Whitney U test.
The resulting $p$-values were then corrected as well using the Benjamini–Hochberg procedure, with differences at $q < 0.05$ interpreted as statistically significant.

For the graphical representation of strategy switching tendencies, an aggregated STM was additionally created by summing the transition count matrices $\sum_s M_s$ across all sessions and subsequently converting them row-wise into probabilities.
This representation provides an intuitively interpretable overview of switching structures.
The significances previously determined at the session level were added as an overlay in the form of markers, thereby combining visual clarity and inferential statistical support within a single figure; values with $q < 0.05$ are indicated by an asterisk (*) to denote statistical significance, while values approaching significance ($q < 0.10$) are indicated by a dagger (†) to highlight observable trends.
To further enhance interpretability, the switching patterns were additionally consolidated at the level of the three strategy groups (\textit{exploratory problem-solving}, \textit{structured problem-solving}, \textit{non-progressive behavior}) in order to make shifts in transition dynamics at the group level  explicitly visible.
\begin{figure*}[!ht]
  \centering
  \newlength{\rest}
  \setlength{\rest}{\dimexpr\textwidth-.5cm\relax}
  \begin{subfigure}[b]{\dimexpr 7\rest/19\relax}
    \centering
    \includegraphics[width=\linewidth]{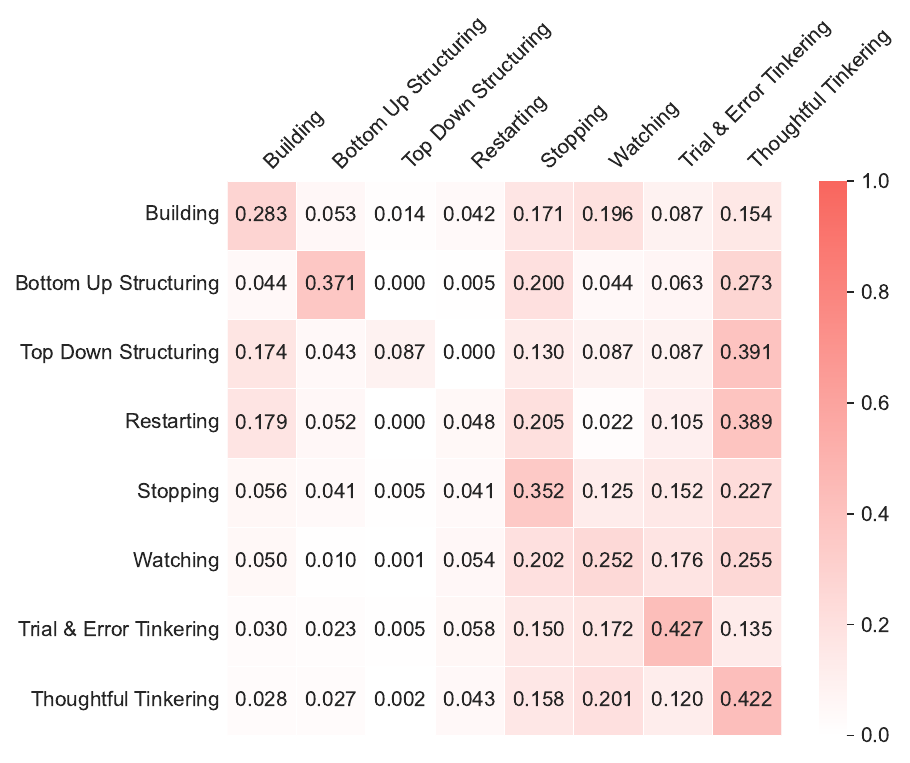}
  \end{subfigure}\hspace{0.25cm}%
  \begin{subfigure}[b]{\dimexpr 7\rest/19\relax}
    \centering
    \includegraphics[width=\linewidth]{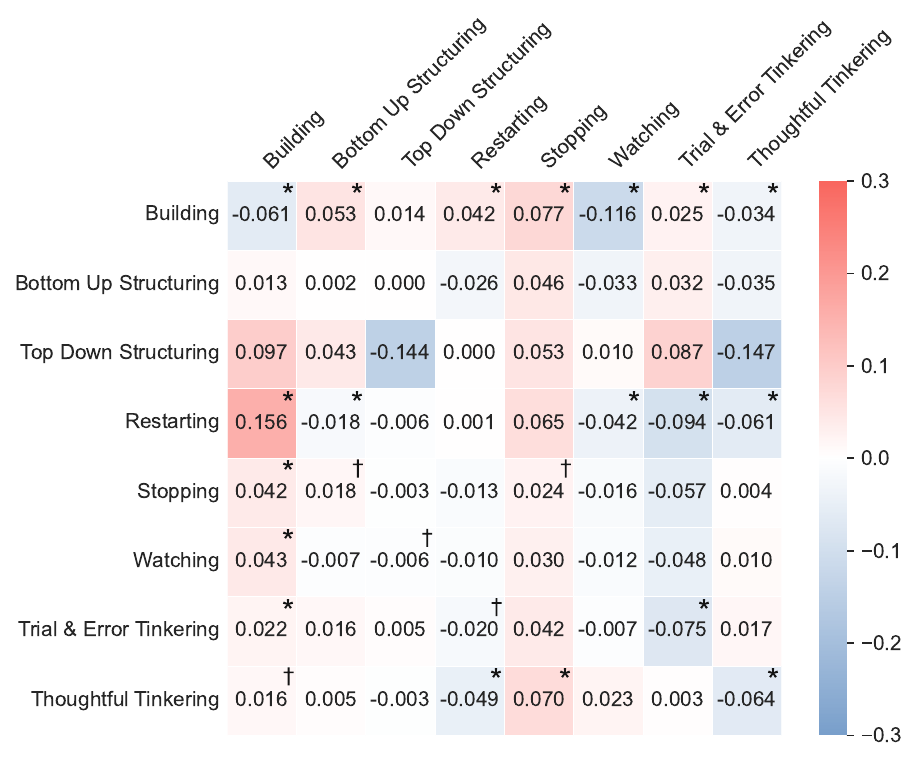}
  \end{subfigure}\hspace{0.25cm}%
  \begin{subfigure}[b]{\dimexpr 5\rest/19\relax}
    \centering
    \includegraphics[width=\linewidth]{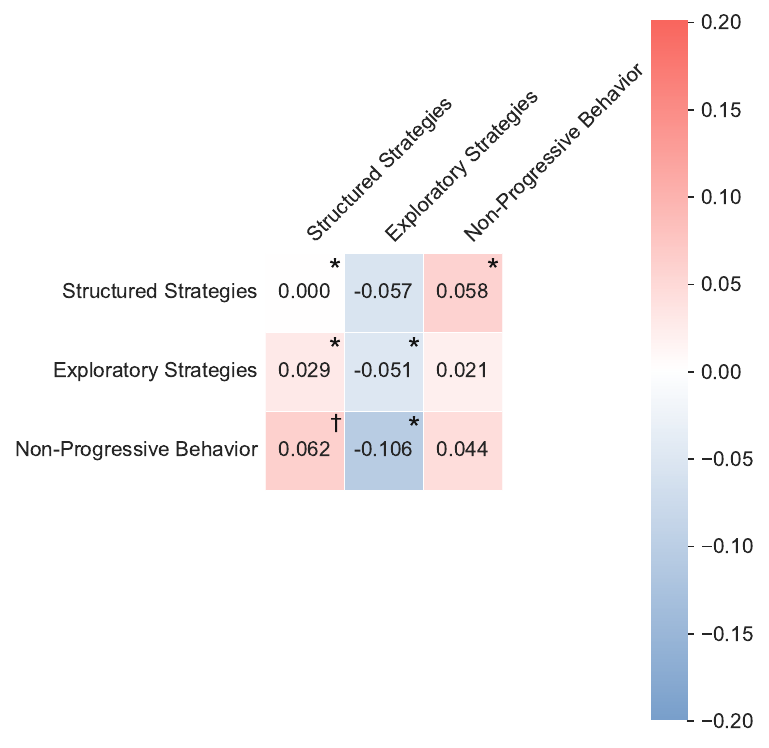}
  \end{subfigure}%
  
  \caption{Strategy Transition Matrices illustrating transition probabilities from row to column (left), full differences (middle), and aggregated strategy group differences (right) between the baseline and feedback game versions ($*$ $q < 0.05$, † $q < 0.10$)}
  
  \Description{Strategy Transition Matrices illustrating transition probabilities from row to column, full differences, and aggregated strategy group differences between the baseline and feedback game versions}
  \label{fig:stm}
\end{figure*}

\section{Results}
Of the total of 230 recorded gameplay sessions, 205 sessions were included in the analysis after systematic data cleaning.
The main reasons for the exclusion of individual sessions were suspected connection interruptions between the game and feedback tablets as well as very short tutorial durations and playtimes.
The coded data, analysis scripts, and supplementary materials are openly available via the OSF repository \cite{OSFRepository} to ensure transparency and reproducibility.
Among the included sessions, 52 learners (25.3\%) reached the game objective, representing a slight decline compared to the baseline study (31.5\%).
The average playtime was 14:23 minutes, with a statistically significant difference between successful (13:16) and unsuccessful sessions (14:46) ($p=0.009$).

\subsection{Distribution of Problem-Solving Strategies} 
As in the baseline study, the group of exploratory problem-solving strategies (\textit{Thoughtful Tinkering}, \textit{Trial \& Error Tinkering}, \textit{Watching}) was predominant across gameplay sessions, accounting for a total of 63.95\% of all observation intervals.
In comparison, the shares of \textit{Trial \& Error Tinkering} (-5.53\%) and \textit{Thoughtful Tinkering} (-3.04\%) decreased, while the proportion of the \textit{Watching} strategy (-0.16\%) remained nearly unchanged.
In contrast, in the group of structured problem-solving strategies increases were observed for \textit{Building} (+5.18\%*) and \textit{Bottom-Up Structuring} (+1.38\%), while the proportion of \textit{Top-Down Structuring} (-0.13\%) remained nearly identical.
Overall, the balance thus shifted in favor of the structured strategies group (+6.43\%*) and at the expense of the exploratory strategies group (-8.73\%*).
The average proportion of structured problem-solving strategies increased in both successful (23.74\%, +13.16\%*) and in unsuccessful sessions (6.57\%, +4.12\%*).
Furthermore, complete restructurings of solution approaches in the sense of the \textit{Restarting} pattern showed a slight reduction of 2.65\%*, while the share of playtime without relevant problem-solving activity (\textit{Stopping}) exhibited a more pronounced increase of 4.95\%*.
A detailed temporal analysis of the \textit{Stopping} proportion in observation intervals in which new feedback messages were presented revealed that \textit{Stopping} occurred particularly frequently after the first feedback message ($\approx$29\%), but subsequently declined to a markedly lower level by the sixth message ($\approx$12–15\%).
A trend analysis across the first ten feedback messages confirmed this decline statistically ($\rho = -0.915, p < 0.001$).
Later fluctuations in the \textit{Stopping} proportion must be interpreted with caution, owing to the reduced number of sessions at higher feedback message counts.

\subsection{Correlations with Game Success}
In the baseline study, the relative shares of the structured problem-solving strategies group correlated most strongly and positively with gameplay success ($\rho = 0.59, p < 0.001$), whereas those of the exploratory strategies group showed the strongest negative correlation ($\rho = -0.41, p < 0.001$).
In the data of the feedback version, a more nuanced pattern emerged: here, the combined shares of \textit{Bottom-Up Structuring} and \textit{Building} ($\rho = 0.627, p < 0.001$) exhibited the strongest positive correlation, whereas the shares of the group consisting of \textit{Restarting}, \textit{Stopping}, \textit{Trial \& Error Tinkering}, and \textit{Watching} showed the strongest negative correlation ($\rho = -0.586, p < 0.001$).

\subsection{Strategy Switching Behavior}
The strategy transition matrices (see Fig. \ref{fig:stm}) present the observed strategy transitions across consecutive 30-second observation intervals. 
As in the baseline study, persistence within strategies dominated overall. However, significant reductions in stability were observed for \textit{Trial \& Error Tinkering} (-7.5\%*), and \textit{Thoughtful Tinkering} (-6.4\%*). 
Accordingly, at the aggregated level of strategy groups, the analysis revealed a significant decline in the persistence of exploratory strategies (-5.1\%*), while transitions into structured strategies (+2.9\%*) and non-progressive behavior (+2.1\%) increased. 
In the baseline study, restarts often transitioned to tinkering strategies, but in the present data these pathways decreased for both \textit{Thoughtful Tinkering} (–6.1\%*) and \textit{Trial \& Error Tinkering} (–9.4\%*).
Conversely, restarts were much more often followed directly by \textit{Building} (+15.6\%*). 
Beyond restarts, direct transitions into \textit{Building} increased significantly from \textit{Stopping} (+4.2\%*) and from \textit{Watching} (+4.3\%*). 
\section{Discussion}
The present study investigated the effectiveness of an adaptive, strategy-oriented feedback system on the structuring of students’ problem-solving processes in the development of decision trees within a self-organized, game-based learning setting.
The primary design emphasis of this feedback system was not placed on the correction or optimization of learning outcomes, but rather on the targeted modulation of learners’ problem-solving strategies and methodological approaches during the ongoing enactment of the task.
In comparison to the baseline study without an adaptive feedback system, the data from the current investigation reveal a clear shift in the behavioral repertoire of the learners.
Structured problem-solving strategies occurred more frequently overall, while exploratory approaches exhibited decreased persistence.
Furthermore, phases of non-progressive behavior (\textit{Stopping}, \textit{Restarting}) were more often converted into structured approaches, while transitions into exploratory problem-solving strategies decreased.
The frequency of transitions within structured strategies, by contrast, remained largely stable.
This suggests that the developed feedback system primarily facilitates the initiation of structured approaches, while contributing to their long-term stabilization only to a limited extent.
Accordingly, it can be stated that the feedback system induces short-term procedural changes in problem-solving behavior by prompting learners to transition more frequently from exploratory to structured strategies.
However, these improvements in problem-solving procedure did not immediately translate into a higher success rate within the game.
It appears plausible to assume that the procedural gains only materialize as durable learning outcomes after repeated application and consolidation.
The fine-grained temporal analysis of \textit{Stopping} phases in relation to newly presented feedback messages suggests that learners progressively allocated less attention to the feedback as the game progressed.
This decline in attentional engagement may be attributable to several factors: On the one hand, the feedback messages might have been perceived as less informative once learners had internalized the core game mechanics and gained routine. On the other hand, it is plausible that learners skimmed the later feedback messages rather than processing them in depth, thereby obviating the need for extended pauses to extract the relevant information.
Overall, it can be concluded that the feedback system demonstrated effectiveness as a catalyst for shifts toward structured problem-solving approaches, although additional scaffolding appears necessary to ensure their sustained consolidation.

From a didactic perspective, this set of findings is noteworthy in several respects.
The observed transition patterns indicate that the feedback system particularly promotes transitions from exploratory or non-progressive phases into structured problem-solving strategies.
In the context of learning tasks related to ML, this is of particular importance, as problem-solving in this domain, unlike in traditional computer science tasks, is predominantly inductive, data-driven, and shaped by heuristic reasoning.
By supporting precisely these initial transitions, the feedback system addresses a point at which ML-specific problem-solving practices differ substantially from established routines in computing education, thereby helping to prepare learners for the particularities of data-driven modeling processes.
At the same time, it must be taken into account that the transitions triggered by the feedback system may entail additional cognitive load for the learners.
The demand to abandon exploratory behavior and shift toward structured approaches requires the reorganization of existing routines and the internalization of new strategies, whose consolidation necessitates repeated application.
During this transitional phase, the efficiency of problem-solving may temporarily decline, even though the quality of the approach already improves.
Against this backdrop, it appears plausible that the procedural gains are clearly reflected in the observed strategic patterns, but only translate to stable performance outcomes through repeated application and consolidating learning opportunities.

These observations yield implications for classroom practices aimed at fostering problem-solving competencies in the field of ML.
Feedback is particularly effective when it specifically targets phases of strategy change and encourages learners to engage in explicit planning.
In the development of decision trees, this could be operationalized didactically through routines that, before every restart or change in problem-solving direction, require learners to articulate a brief planning rule; for example, which model modification or adaptation should be examined next.
Similarly, passive or observation phases can be transformed into evidence-based decisions through short, strategy-oriented questions.
Transparent labeling of the strategy currently being pursued could further support these processes by facilitating learners’ reflection on their own approaches.
These implications thus provide a metacognitive-instructional framework for the design of further teaching and learning arrangements, the empirical examination of which should be systematically pursued in future research.
\section{Limitations}
While the adaptive, strategy-oriented feedback effectively triggered short-term transitions from exploratory to structured problem-solving strategies, these behaviors did not consolidate over time. 
From a learning-scientific perspective, this phenomenon can be explained by several interrelated mechanisms. 
Firstly, feedback-induced strategy shifts likely increased extraneous cognitive load \cite{swellerCognitiveArchitectureInstructional2019}, as learners simultaneously had to process task-related and metacognitive information. 
This dual demand may have exceeded the capacity of working memory, thereby diminishing the cognitive resources available for the consolidation of new strategies into long-term memory. 
Secondly, from the perspective of self-regulated learning \cite{winneStudyingSelfregulatedLearning1998}, the implemented feedback primarily supported metacognitive monitoring but provided limited scaffolding for metacognitive control. 
In the absence of explicit prompts for planning, goal setting, and evaluation, learners may temporarily adopt structured strategies but fail to maintain them once external regulation diminishes. 
Thirdly, as stated in Feedback Intervention Theory \cite{klugerEffectsFeedbackInterventions1996}, the attentional focus elicited by feedback tends to diminish over time. 
As learners' task schemas become more automated, subsequent feedback messages are often processed only superficially, thereby reducing opportunities for reflection and deep learning. Consequently, motivational dynamics may have exerted a constraining influence on persistence. 
According to the principles of control-value theory \cite{pekrunControlValueTheoryAchievement2006}, the provision of feedback that challenges established routines without prompting immediate success experiences can lead to a reduction in perceived competence and an escalation in frustration. 
This, in turn, can result in a reversion to less demanding exploratory strategies. Taken together, the findings of the present study indicate that while the investigated feedback system successfully initiated cognitive restructuring, it did not yet optimize the conditions for the sustained consolidation of problem-solving strategies. 
Consequently, future research should incorporate mechanisms for cognitive load management, explicit metacognitive control, and motivational regulation. 
Examples of such mechanisms include faded feedback, learner-generated planning prompts, and reflective summaries. 
The integration of these mechanisms is expected to promote the durable internalization of structured problem-solving behavior.

\section{Outlook}
This study provides empirical evidence that an adaptive, strategy-oriented feedback system can effectively support learners in the self-regulated development of decision trees within the digital learning game \textit{Match the Monkeys}, thereby fostering the adoption of structured problem-solving strategies.
These effects were particularly salient in the transitions observed from unproductive or exploratory problem-solving phases to more systematic approaches.
Building on these findings, future research should examine how the observed procedural gains can be sustained and consolidated over time through targeted refinements of the feedback design.
It is likewise necessary to investigate whether comparable effects can be replicated in other ML-related learning tasks and learning environments, in order to strengthen external validity and generalizability.
In addition, it appears promising to enhance the validity of strategy detection through the integration of multiple behavioral and process-oriented data sources and to systematically assess learners’ awareness and effective use of the feedback provided.
The results obtained in the game-based context can thus serve as a foundation for the curricular advancement and further development of learning environments in ML education.
By integrating strategy-oriented feedback systematically into digital learning settings, opportunities arise to sustainably strengthen the development of structured problem-solving competencies in the fields of AI and ML.

\bibliographystyle{ACM-Reference-Format}
\bibliography{bibliography}

@misc{OSFRepository,
  author = {Witt, Clemens and Leonhardt, Thiemo and Marx, Erik and Grillenberger, Mareen},
  title = {OSF Repository: Strategy-Oriented Feedback for Fostering Systematic Problem-Solving in Machine Learning Education},
  doi = {10.17605/OSF.IO/NC3DM},
  year  = {2025},
}

@misc{brennanNewFrameworksStudying2012,
  title = {New Frameworks for Studying and Assessing the Development of Computational Thinking},
  author = {Brennan, Karen and Resnick, Mitchel},
  year = {2012},
  url = {https://scratched.gse.harvard.edu/ct/files/AERA2012.pdf},
  langid = {english}
}

@inproceedings{conwayHCAIBlockModel2024,
  title = {{{HCAI Block Model}}: {{A}} Competence Model for {{Human Centred Artificial Intelligence}} at {{K-12}}},
  shorttitle = {{{HCAI Block Model}}},
  booktitle = {Proceedings of the 2024 {{Conference}} on {{Human Centred Artificial Intelligence}} - {{Education}} and {{Practice}}},
  author = {Conway, Brian and Nolan, Keith and Quille, Keith},
  year = {2024},
  month = dec,
  pages = {22--28},
  publisher = {ACM},
  address = {Naples Italy},
  doi = {10.1145/3701268.3701273},
  urldate = {2025-09-03},
  isbn = {979-8-4007-1159-6},
  langid = {english}
}

@article{fleischerTEACHINGLEARNINGCONSTRUCT2024,
  title = {{{Teaching and learning to construct data-based decision trees using data cards as the first introduction to machine learning in middle school}}},
  author = {Fleischer, Yannik and Podworny, Susanne and Biehler, Rolf},
  year = {2024},
  month = aug,
  journal = {STATISTICS EDUCATION RESEARCH JOURNAL},
  volume = {23},
  number = {1},
  pages = {3},
  issn = {1570-1824},
  doi = {10.52041/serj.v23i1.450},
  urldate = {2025-09-03}
}

@article{groverComputationalThinkingCompetency2018,
  title = {Computational Thinking: {{A}} Competency Whose Time Has Come},
  shorttitle = {Computational Thinking},
  author = {Grover, Shuchi and Pea, Roy},
  year = {2018},
  journal = {Computer science education: Perspectives on teaching and learning in school},
  volume = {19},
  number = {1},
  pages = {19--38}
}

@article{groverComputationalThinkingK122013,
  title = {Computational {{Thinking}} in {{K}}--12: {{A Review}} of the {{State}} of the {{Field}}},
  shorttitle = {Computational {{Thinking}} in {{K}}--12},
  author = {Grover, Shuchi and Pea, Roy},
  year = {2013},
  month = jan,
  journal = {Educational Researcher},
  volume = {42},
  number = {1},
  pages = {38--43},
  publisher = {American Educational Research Association},
  issn = {0013-189X},
  doi = {10.3102/0013189X12463051},
  urldate = {2025-09-01}
}

@inbook{hazzanProblemSolvingStrategies2020,
  title = {Problem-{{Solving Strategies}}},
  booktitle = {Guide to {{Teaching Computer Science}}},
  author = {Hazzan, Orit and Ragonis, Noa and Lapidot, Tami},
  year = {2020},
  pages = {143--168},
  publisher = {Springer International Publishing},
  address = {Cham},
  doi = {10.1007/978-3-030-39360-1_8},
  urldate = {2025-08-31},
  collaborator = {Hazzan, Orit and Ragonis, Noa and Lapidot, Tami},
  isbn = {978-3-030-39359-5 978-3-030-39360-1},
  langid = {english}
}

@article{kimSystematicReviewEvaluation2025,
  title = {A Systematic Review of the Evaluation in {{K-12}} Artificial Intelligence Education from 2013 to 2022},
  author = {Kim, Keunjae and Kwon, Kyungbin},
  year = {2025},
  month = jan,
  journal = {Interactive Learning Environments},
  volume = {33},
  number = {1},
  pages = {103--131},
  publisher = {Routledge},
  issn = {1049-4820},
  doi = {10.1080/10494820.2024.2335499},
  urldate = {2025-09-03}
}

@article{liuReimaginingMachineLearning2023,
  title = {Reimagining the Machine Learning Life Cycle to Improve Educational Outcomes of Students},
  author = {Liu, Lydia T. and Wang, Serena and Britton, Tolani and Abebe, Rediet},
  year = {2023},
  month = feb,
  journal = {Proceedings of the National Academy of Sciences},
  volume = {120},
  number = {9},
  pages = {e2204781120},
  publisher = {Proceedings of the National Academy of Sciences},
  doi = {10.1073/pnas.2204781120},
  urldate = {2025-09-02}
}

@article{maharaniPROBLEMSOLVINGCONTEXT2019,
  title = {{{Problem Solving in the Context of Computational Thinking}}},
  author = {Maharani, Swasti and Kholid, Muhammad Noor and Pradana, Lingga Nico and Nusantara, Toto},
  year = {2019},
  month = sep,
  journal = {Infinity Journal},
  volume = {8},
  number = {2},
  pages = {109},
  issn = {2460-9285, 2089-6867},
  doi = {10.22460/infinity.v8i2.p109-116},
  urldate = {2025-08-31},
  copyright = {http://creativecommons.org/licenses/by-sa/4.0},
  langid = {english}
}

@inproceedings{morales-navarroConstructionistApproachesLearning2023,
   title = {Constructionist approaches to learning artificial intelligence/machine learning: Past, present, and future},
   author = {Morales-Navarro, Luis and Kafai, Yasmin B. and Khan, Ken and Romeike, Ralf and Michaeli, Tilman and DiPaola, Daniella and Ali, Safinah and Williams, Randi and Breazeal, Cynthia and Castro, Francisco and DesPortes, Kayla and Stager, Gary and Kumar, Vishesh and Bodon, Herminio and Worsley, Marcelo and Lee, Victor R. and Sarin, Parth and Xie, Benjamin and Wolf, Jacob and Sieh, Isabel and Dennison, Deepak Varuvel and Garcia, Raycelle and Solomon, Cynthia},
   year = {2023},
   publisher = {ETC Press},
   booktitle = {Proceedings of Constructionism 2023},
   pages = {245--254},
   isbn = {978-1-300-99191-5},
   address = {Pittsburgh},
   url = {https://aprendizagemcriativa.org/sites/default/files/2024-12/confab2023.101424.pdf}
}

@inproceedings{morales-navarroUnpackingApproachesLearning2024,
  title = {Unpacking {{Approaches}} to {{Learning}} and {{Teaching Machine Learning}} in {{K-12 Education}}: {{Transparency}}, {{Ethics}}, and {{Design Activities}}},
  shorttitle = {Unpacking {{Approaches}} to {{Learning}} and {{Teaching Machine Learning}} in {{K-12 Education}}},
  booktitle = {Proceedings of the 19th {{WiPSCE Conference}} on {{Primary}} and {{Secondary Computing Education Research}}},
  author = {{Morales-Navarro}, Luis and Kafai, Yasmin B},
  year = {2024},
  month = sep,
  pages = {1--10},
  publisher = {ACM},
  address = {Munich Germany},
  doi = {10.1145/3677619.3678117},
  urldate = {2025-09-02},
  isbn = {979-8-4007-1005-6},
  langid = {english}
}

@inproceedings{nowackModelbasedThinkingPractice2014,
  title = {Model-Based Thinking and Practice: A Top-down Approach to Computational Thinking},
  shorttitle = {Model-Based Thinking and Practice},
  booktitle = {Proceedings of the 14th {{Koli Calling International Conference}} on {{Computing Education Research}}},
  author = {Nowack, Palle and Caspersen, Michael E.},
  year = {2014},
  month = nov,
  pages = {147--151},
  publisher = {ACM},
  address = {Koli Finland},
  doi = {10.1145/2674683.2674686},
  urldate = {2025-09-02},
  isbn = {978-1-4503-3065-7},
  langid = {english}
}

@inproceedings{olariDatarelatedPracticesCreating2024a,
  title = {Data-Related Practices for Creating {{Artificial Intelligence}} Systems in {{K-12}}},
  booktitle = {Proceedings of the 19th {{WiPSCE Conference}} on {{Primary}} and {{Secondary Computing Education Research}}},
  author = {Olari, Viktoriya and Romeike, Ralf},
  year = {2024},
  month = sep,
  pages = {1--10},
  publisher = {ACM},
  address = {Munich Germany},
  doi = {10.1145/3677619.3678115},
  urldate = {2025-09-03},
  isbn = {979-8-4007-1005-6},
  langid = {english}
}

@misc{papertConstructionismNewOpportunity1986,
  title = {Constructionism: {{A}} New Opportunity for Elementary Science Education},
  author = {Papert, Seymour},
  year = {1986},
  note = {A Proposal to The National Science Foundation},
  publisher = {{Massachusetts Institute of Technology, Media Laboratory, Epistemology and Learning Group}}
}

@inproceedings{passiProblemFormulationFairness2019,
  title = {Problem {{Formulation}} and {{Fairness}}},
  booktitle = {Proceedings of the {{Conference}} on {{Fairness}}, {{Accountability}}, and {{Transparency}}},
  author = {Passi, Samir and Barocas, Solon},
  year = {2019},
  month = jan,
  pages = {39--48},
  publisher = {ACM},
  address = {Atlanta GA USA},
  doi = {10.1145/3287560.3287567},
  urldate = {2025-09-02},
  isbn = {978-1-4503-6125-5},
  langid = {english}
}

@book{polyaHowSolveIt1945,
  title = {How to {{Solve It}}: {{A New Aspect}} of {{Mathematical Method}}},
  author = {Polya, George.},
  year = {1945},
  month = dec,
  publisher = {Princeton University Press},
  address = {Princeton},
  doi = {10.1515/9781400828678},
  urldate = {2025-09-04},
  isbn = {978-1-4008-2867-8}
}

@article{shuteDemystifyingComputationalThinking2017,
  title = {Demystifying Computational Thinking},
  author = {Shute, Valerie J. and Sun, Chen and {Asbell-Clarke}, Jodi},
  year = {2017},
  month = nov,
  journal = {Educational Research Review},
  volume = {22},
  pages = {142--158},
  issn = {1747938X},
  doi = {10.1016/j.edurev.2017.09.003},
  urldate = {2025-09-01},
  langid = {english}
}

@inproceedings{tedreCT202021a,
  title = {{{CT}} 2.0},
  booktitle = {Proceedings of the 21st {{Koli Calling International Conference}} on {{Computing Education Research}}},
  author = {Tedre, Matti and Denning, Peter and Toivonen, Tapani},
  year = {2021},
  month = nov,
  pages = {1--8},
  publisher = {ACM},
  address = {Joensuu Finland},
  doi = {10.1145/3488042.3488053},
  urldate = {2025-08-31},
  isbn = {978-1-4503-8488-9},
  langid = {english}
}

@article{weintropDefiningComputationalThinking2016,
  title = {Defining {{Computational Thinking}} for {{Mathematics}} and {{Science Classrooms}}},
  author = {Weintrop, David and Beheshti, Elham and Horn, Michael and Orton, Kai and Jona, Kemi and Trouille, Laura and Wilensky, Uri},
  year = {2016},
  journal = {Journal of Science Education and Technology},
  volume = {25},
  number = {1},
  eprint = {43867736},
  eprinttype = {jstor},
  pages = {127--147},
  publisher = {Springer},
  issn = {1059-0145},
  urldate = {2025-09-02}
}

@article{wingComputationalThinking2006a,
  title = {Computational Thinking},
  author = {Wing, Jeannette M.},
  year = {2006},
  month = mar,
  journal = {Communications of the ACM},
  volume = {49},
  number = {3},
  pages = {33--35},
  issn = {0001-0782, 1557-7317},
  doi = {10.1145/1118178.1118215},
  urldate = {2025-09-01},
  langid = {english}
}

@article{wingComputationalThinkingThinking2008,
  title = {Computational Thinking and Thinking about Computing},
  author = {Wing, Jeannette M.},
  year = {2008},
  month = oct,
  journal = {Philosophical transactions. Series A, Mathematical, physical, and engineering sciences},
  volume = {366},
  number = {1881},
  pages = {3717--3725},
  issn = {1364-503X},
  doi = {10.1098/rsta.2008.0118},
  urldate = {2025-09-01},
  pmcid = {PMC2696102},
  pmid = {18672462}
}

@article{wooProblemSolvedHow2022b,
  title = {Problem Solved, but How? {{An}} Exploratory Study into Students' Problem Solving Processes in Creative Coding Tasks},
  shorttitle = {Problem Solved, but How?},
  author = {Woo, Karen and Falloon, Garry},
  year = {2022},
  month = dec,
  journal = {Thinking Skills and Creativity},
  volume = {46},
  pages = {101193},
  issn = {18711871},
  doi = {10.1016/j.tsc.2022.101193},
  urldate = {2025-08-31},
  langid = {english}
}

@book{zimmermanTeachingAIExploring2018,
  title = {{Teaching AI: Exploring New Frontiers for Learning}},
  shorttitle = {{Teaching AI}},
  author = {Zimmerman, Michelle},
  year = {2018},
  month = dec,
  publisher = {Internation Society for Technology in Education},
  address = {Portland, Oregon},
  isbn = {978-1-56484-705-8},
  langid = {Englisch}
}

@inproceedings{akramImprovingStealthAssessment,
  title={Improving stealth assessment in game-based learning with LSTM-based analytics},
  author={Akram, Bita and Min, Wookhee and Wiebe, Eric and Mott, Bradford and Boyer, Kristy Elizabeth and Lester, James},
  booktitle={International conference on educational data mining},
  year={2018}
}

@incollection{dicerboFutureAssessmentTechnologyRich2016,
  title = {The {{Future}} of {{Assessment}} in {{Technology-Rich Environments}}: {{Psychometric Considerations}}},
  shorttitle = {The {{Future}} of {{Assessment}} in {{Technology-Rich Environments}}},
  booktitle = {Learning, {{Design}}, and {{Technology}}},
  author = {DiCerbo, Kristen E and Shute, Valerie and Kim, Yoon Jeon},
  editor = {Spector, Michael J and Lockee, Barbara B and Childress, Marcus D.},
  year = {2016},
  pages = {1--21},
  publisher = {Springer International Publishing},
  address = {Cham},
  doi = {10.1007/978-3-319-17727-4_66-1},
  urldate = {2025-04-03},
  isbn = {978-3-319-17727-4},
  langid = {english}
}

@article{emersonMultimodalLearningAnalytics2020a,
  title = {Multimodal Learning Analytics for Game-Based Learning},
  author = {Emerson, Andrew and Cloude, Elizabeth B. and Azevedo, Roger and Lester, James},
  year = {2020},
  journal = {British Journal of Educational Technology},
  volume = {51},
  number = {5},
  pages = {1505--1526},
  issn = {1467-8535},
  doi = {10.1111/bjet.12992},
  urldate = {2025-04-04},
  copyright = {{\copyright} 2020 British Educational Research Association},
  langid = {english}
}

@book{fruhInhaltsanalyse2017,
  title = {Inhaltsanalyse},
  author = {Fr{\"u}h, Werner},
  edition = {9},
  year = {2017},
  month = feb,
  series = {Utb-{{Titel}} Ohne {{Reihe}}},
  publisher = {UVK Verlag},
  address = {Konstanz and München},
  doi = {10.36198/9783838547350},
  urldate = {2025-04-06},
  isbn = {978-3-8252-4735-5}
}

@article{guptaMultimodalMultiTaskStealth2021,
  title = {Multimodal {{Multi-Task Stealth Assessment}} for {{Reflection-Enriched Game-Based Learning}}},
  author = {Gupta, Anisha and Carpenter, Dan and Min, Wookhee and Rowe, Jonathan and Azevedo, Roger and Lester, James},
  year = {2021},
  langid = {english}
}

@article{hendersonEnhancingStealthAssessment2022,
  title = {Enhancing {{Stealth Assessment}} in {{Game-Based Learning Environments}} with {{Generative Zero-Shot Learning}}},
  author = {Henderson, Nathan and Acosta, Halim and {Wookhee Min} and Mott, Bradford and Trudi Lord and Reichsman, Frieda and Dorsey, Chad and Wiebe, Eric and Lester, James},
  editor = {Mitrovic, Antonija and Bosch, Nigel},
  year = {2022},
  month = jul,
  publisher = {Zenodo},
  doi = {10.5281/ZENODO.6852942},
  urldate = {2025-04-03},
  copyright = {Creative Commons Attribution 4.0 International, Open Access},
  langid = {english}
}

@inproceedings{lindnerUnpluggedActivitiesContext2019,
  title = {Unplugged {{Activities}} in the {{Context}} of {{AI}}},
  booktitle = {Informatics in {{Schools}}. {{New Ideas}} in {{School Informatics}}: 12th {{International Conference}} on {{Informatics}} in {{Schools}}: {{Situation}}, {{Evolution}}, and {{Perspectives}}, {{ISSEP}} 2019, {{Larnaca}}, {{Cyprus}}, {{November}} 18--20, 2019, {{Proceedings}}},
  author = {Lindner, Annabel and Seegerer, Stefan and Romeike, Ralf},
  year = {2019},
  month = nov,
  pages = {123--135},
  publisher = {Springer-Verlag},
  address = {Berlin, Heidelberg},
  doi = {10.1007/978-3-030-33759-9_10},
  urldate = {2024-05-18},
  isbn = {978-3-030-33758-2}
}

@article{minDeepStealthGameBasedLearning2020,
  title = {{{DeepStealth}}: {{Game-Based Learning Stealth Assessment With Deep Neural Networks}}},
  shorttitle = {{{DeepStealth}}},
  author = {Min, Wookhee and Frankosky, Megan H. and Mott, Bradford W. and Rowe, Jonathan P. and Smith, Andy and Wiebe, Eric and Boyer, Kristy Elizabeth and Lester, James C.},
  year = {2020},
  month = apr,
  journal = {IEEE Transactions on Learning Technologies},
  volume = {13},
  number = {2},
  pages = {312--325},
  issn = {1939-1382},
  doi = {10.1109/TLT.2019.2922356},
  urldate = {2025-01-29}
}

@article{mislevyDesignDiscoveryEducational2012,
  title = {Design and {{Discovery}} in {{Educational Assessment}}: {{Evidence-Centered Design}}, {{Psychometrics}}, and {{Educational Data Mining}}},
  shorttitle = {Design and {{Discovery}} in {{Educational Assessment}}},
  author = {Mislevy, Robert J. and Behrens, John T. and Dicerbo, Kristen E. and Levy, Roy},
  year = {2012},
  month = oct,
  journal = {Journal of Educational Data Mining},
  volume = {4},
  number = {1},
  pages = {11--48},
  doi = {10.5281/zenodo.3554642},
  urldate = {2025-04-03},
  langid = {english}
}

@article{mislevyFocusArticleStructure2003,
  title = {Focus {{Article}}: {{On}} the {{Structure}} of {{Educational Assessments}}},
  shorttitle = {Focus {{Article}}},
  author = {Mislevy, Robert J. and Steinberg, Linda S. and Almond, Russell G.},
  year = {2003},
  month = jan,
  journal = {Measurement: Interdisciplinary Research \& Perspective},
  volume = {1},
  number = {1},
  pages = {3--62},
  issn = {1536-6367, 1536-6359},
  doi = {10.1207/S15366359MEA0101_02},
  urldate = {2025-04-03},
  langid = {english}
}

@article{shuteSimplyAssessment2009,
  title = {Simply Assessment},
  author = {Shute, Valerie J.},
  year = {2009},
  journal = {International Journal of Learning and Media},
  volume = {1},
  number = {2},
  pages = {1--11},
  publisher = {MIT Press},
  urldate = {2025-04-03}
}

@book{shuteStealthAssessmentMeasuring2013,
  title = {Stealth {{Assessment}}: {{Measuring}} and {{Supporting Learning}} in {{Video Games}}},
  shorttitle = {Stealth {{Assessment}}},
  author = {Shute, Valerie and Ventura, Matthew},
  year = {2013},
  month = mar,
  publisher = {The MIT Press},
  doi = {10.7551/mitpress/9589.001.0001},
  address = {Cambridge, MA and London},
  urldate = {2025-04-03},
  copyright = {https://creativecommons.org/licenses/by-nc-nd/4.0/},
  isbn = {978-0-262-31521-0},
  langid = {english}
}

@incollection{wittGuessworkGamePlan2024a,
  title = {From {{Guesswork}} to {{Game Plan}}: {{Exploring Problem-Solving-Strategies}} in a {{Machine Learning Game}}},
  shorttitle = {From {{Guesswork}} to {{Game Plan}}},
  booktitle = {Informatics in {{Schools}}. {{Innovative Approaches}} to {{Computer Science Teaching}} and {{Learning}}},
  author = {Witt, Clemens and Leonhardt, Thiemo and Marx, Erik and Bergner, Nadine},
  editor = {Pluh{\'a}r, Zsuzsa and Ga{\'a}l, Bence},
  year = {2024},
  volume = {15228},
  pages = {73--84},
  publisher = {Springer Nature Switzerland},
  address = {Cham},
  doi = {10.1007/978-3-031-73474-8_6},
  urldate = {2024-10-17},
  isbn = {978-3-031-73473-1 978-3-031-73474-8},
  langid = {english}
}

@inproceedings{wittMultimodalLateFusion2026,
  title = {Multimodal {{Late Fusion Model}} for~{{Problem-Solving Strategy Classification}} in~a~{{Machine Learning Game}}},
  booktitle = {Two {{Decades}} of {{TEL}}. {{From Lessons Learnt}} to {{Challenges Ahead}}},
  author = {Witt, Clemens and Leonhardt, Thiemo and Bergner, Nadine and Grillenberger, Mareen},
  editor = {Tammets, Kairit and Sosnovsky, Sergey and Ferreira Mello, Rafael and Pishtari, Gerti and Nazaretsky, Tanya},
  year = {2026},
  pages = {281--286},
  publisher = {Springer Nature Switzerland},
  address = {Cham},
  doi = {10.1007/978-3-032-03873-9_37},
  isbn = {978-3-032-03873-9},
  langid = {english}
}

@article{butlerFeedbackSelfRegulatedLearning1995,
  title = {Feedback and {{Self-Regulated Learning}}: {{A Theoretical Synthesis}}},
  shorttitle = {Feedback and {{Self-Regulated Learning}}},
  author = {Butler, Deborah L. and Winne, Philip H.},
  year = {1995},
  month = sep,
  journal = {Review of Educational Research},
  volume = {65},
  number = {3},
  pages = {245--281},
  publisher = {American Educational Research Association},
  issn = {0034-6543},
  doi = {10.3102/00346543065003245},
  urldate = {2025-09-15}
}

@article{hattiePowerFeedback2007a,
  title = {The {{Power}} of {{Feedback}}},
  author = {Hattie, John and Timperley, Helen},
  year = {2007},
  month = mar,
  journal = {Review of Educational Research},
  volume = {77},
  number = {1},
  pages = {81--112},
  publisher = {American Educational Research Association},
  issn = {0034-6543},
  doi = {10.3102/003465430298487},
  urldate = {2025-09-15},
  langid = {english}
}

@incollection{kalyugaEvaluatingManagingCognitive2009,
  title = {Evaluating and {{Managing Cognitive Load}} in {{Games}}},
  booktitle = {Handbook of {{Research}} on {{Effective Electronic Gaming}} in {{Education}}},
  author = {Kalyuga, Slava and Plass, Jan L.},
  year = {2009},
  pages = {719--737},
  publisher = {IGI Global Scientific Publishing},
  doi = {10.4018/978-1-59904-808-6.ch041},
  urldate = {2025-09-15},
  copyright = {Access limited to members},
  isbn = {978-1-59904-808-6},
  langid = {english}
}

@article{koskinenStrengthDirectionDifficulty2023,
  title = {The Strength and Direction of the Difficulty Adaptation Affect Situational Interest in Game-Based Learning},
  author = {Koskinen, Antti and McMullen, Jake and {Hannula-Sormunen}, Minna and Ninaus, Manuel and Kiili, Kristian},
  year = {2023},
  month = mar,
  journal = {Computers \& Education},
  volume = {194},
  pages = {104694},
  issn = {0360-1315},
  doi = {10.1016/j.compedu.2022.104694},
  urldate = {2025-09-16}
}

@article{linnenbrink-garciaAdaptiveMotivationEmotion2016,
  title = {Adaptive {{Motivation}} and {{Emotion}} in {{Education}}: {{Research}} and {{Principles}} for {{Instructional Design}}},
  shorttitle = {Adaptive {{Motivation}} and {{Emotion}} in {{Education}}},
  author = {{Linnenbrink-Garcia}, Lisa and Patall, Erika A. and Pekrun, Reinhard},
  year = {2016},
  month = oct,
  journal = {Policy Insights from the Behavioral and Brain Sciences},
  volume = {3},
  number = {2},
  pages = {228--236},
  publisher = {SAGE Publications},
  issn = {2372-7322},
  doi = {10.1177/2372732216644450},
  urldate = {2025-09-15}
}

@article{maierPersonalizedFeedbackDigital2022a,
  title = {Personalized Feedback in Digital Learning Environments: {{Classification}} Framework and Literature Review},
  shorttitle = {Personalized Feedback in Digital Learning Environments},
  author = {Maier, Uwe and Klotz, Christian},
  year = {2022},
  journal = {Computers and Education: Artificial Intelligence},
  volume = {3},
  pages = {100080},
  issn = {2666920X},
  doi = {10.1016/j.caeai.2022.100080},
  urldate = {2025-09-15},
  langid = {english}
}

@book{morrisonDesigningEffectiveInstruction2019,
  title = {Designing {{Effective Instruction}}},
  author = {Morrison, Gary R. and Ross, Steven J. and Morrison, Jennifer R. and Kalman, Howard K.},
  year = {2019},
  month = mar,
  publisher = {John Wiley \& Sons},
  isbn = {978-1-119-46593-5},
  langid = {english}
}

@incollection{narcissFeedbackStrategiesInteractive2008a,
  title = {Feedback {{Strategies}} for {{Interactive Learning Tasks}}},
  booktitle = {Handbook of {{Research}} on {{Educational Communications}} and {{Technology}}},
  author = {Narciss, Susanne},
  address = {New York, NY},
  year = {2008},
  edition = {3},
  publisher = {Routledge}
}

@article{qinjinjiaAssessingFaithfulnessLLMgenerated2024,
  title = {On {{Assessing}} the {{Faithfulness}} of {{LLM-generated Feedback}} on {{Student Assignments}}},
  author = {Qinjin Jia and Jialin Cui and Ruijie Xi and Chengyuan Liu and Parvez Rashid and Ruochi Li and Edward Gehringer},
  editor = {Benjamin, Paa{\ss}en and Carrie, Demmans Epp},
  year = {2024},
  month = jul,
  publisher = {International Educational Data Mining Society},
  doi = {10.5281/ZENODO.12729868},
  urldate = {2025-09-16},
  copyright = {Creative Commons Attribution 4.0 International},
  langid = {english}
}

@article{steinertHarnessingLargeLanguage2024,
  title = {Harnessing Large Language Models to Develop Research-Based Learning Assistants for Formative Feedback},
  author = {Steinert, Steffen and Avila, Karina E. and Ruzika, Stefan and Kuhn, Jochen and K{\"u}chemann, Stefan},
  year = {2024},
  month = dec,
  journal = {Smart Learning Environments},
  volume = {11},
  number = {1},
  pages = {62},
  issn = {2196-7091},
  doi = {10.1186/s40561-024-00354-1},
  urldate = {2025-09-16}
}

@article{wooDigitalGameBasedLearning2014,
  title = {Digital {{Game-Based Learning Supports Student Motivation}}, {{Cognitive Success}}, and {{Performance Outcomes}}},
  author = {Woo, Jeng-Chung},
  year = {2014},
  journal = {Journal of Educational Technology \& Society},
  volume = {17},
  number = {3},
  eprint = {jeductechsoci.17.3.291},
  eprinttype = {jstor},
  pages = {291--307},
  publisher = {International Forum of Educational Technology \& Society},
  issn = {1176-3647},
  urldate = {2025-09-15}
}

@article{shuteFocusFormativeFeedback2008a,
  title = {Focus on {{Formative Feedback}}},
  author = {Shute, Valerie J.},
  year = {2008},
  month = mar,
  journal = {Review of Educational Research},
  volume = {78},
  number = {1},
  pages = {153--189},
  publisher = {American Educational Research Association},
  issn = {0034-6543},
  doi = {10.3102/0034654307313795},
  urldate = {2025-09-16},
  langid = {english}
}

@inproceedings{longWhatAILiteracy2020,
  title = {What Is {{AI Literacy}}? {{Competencies}} and {{Design Considerations}}},
  shorttitle = {What Is {{AI Literacy}}?},
  booktitle = {Proceedings of the 2020 {{CHI Conference}} on {{Human Factors}} in {{Computing Systems}}},
  author = {Long, Duri and Magerko, Brian},
  year = {2020},
  month = apr,
  series = {{{CHI}} '20},
  pages = {1--16},
  publisher = {Association for Computing Machinery},
  address = {Honolulu HI USA},
  doi = {10.1145/3313831.3376727},
  urldate = {2022-07-24},
  isbn = {978-1-4503-6708-0},
  langid = {english}
}

@article{touretzkyMachineLearningFive2022,
  title = {Machine {{Learning}} and the {{Five Big Ideas}} in {{AI}}},
  author = {Touretzky, David and {Gardner-McCune}, Christina and Seehorn, Deborah},
  year = {2022},
  month = oct,
  journal = {International Journal of Artificial Intelligence in Education},
  volume = {33},
  number = {2},
  pages = {1--34},
  issn = {1560-4306},
  doi = {10.1007/s40593-022-00314-1},
  urldate = {2022-10-30},
  langid = {english}
}

@article{ngArtificialIntelligenceAI2024,
  title = {Artificial Intelligence ({{AI}}) Literacy Education in Secondary Schools: A Review},
  shorttitle = {Artificial Intelligence ({{AI}}) Literacy Education in Secondary Schools},
  author = {Ng, Davy Tsz Kit and Su, Jiahong and Leung, Jac Ka Lok and Chu, Samuel Kai Wah},
  year = {2024},
  month = nov,
  journal = {Interactive Learning Environments},
  volume = {32},
  number = {10},
  pages = {6204--6224},
  issn = {1049-4820, 1744-5191},
  doi = {10.1080/10494820.2023.2255228},
  urldate = {2025-09-26},
  langid = {english}
}

@article{benjaminiControllingFalseDiscovery1995,
  title = {Controlling the {{False Discovery Rate}}: {{A Practical}} and {{Powerful Approach}} to {{Multiple Testing}}},
  shorttitle = {Controlling the {{False Discovery Rate}}},
  author = {Benjamini, Yoav and Hochberg, Yosef},
  year = {1995},
  journal = {Journal of the Royal Statistical Society. Series B (Methodological)},
  volume = {57},
  number = {1},
  eprint = {2346101},
  eprinttype = {jstor},
  pages = {289--300},
  publisher = {[Royal Statistical Society, Oxford University Press]},
  issn = {0035-9246},
  urldate = {2025-09-29}
}

@misc{openai2025gpt41,
  author = {OpenAI},
  title = {Introducing GPT-4.1 in the API},
  howpublished = {\url{https://openai.com/index/gpt-4-1/}},
  year = {2025},
  month = apr,
  day = {14},
}

@article{klugerEffectsFeedbackInterventions1996,
  title = {The Effects of Feedback Interventions on Performance: {{A}} Historical Review, a Meta-Analysis, and a Preliminary Feedback Intervention Theory},
  shorttitle = {The Effects of Feedback Interventions on Performance},
  author = {Kluger, Avraham N. and DeNisi, Angelo},
  year = {1996},
  journal = {Psychological Bulletin},
  volume = {119},
  number = {2},
  pages = {254--284},
  publisher = {American Psychological Association},
  address = {US},
  issn = {1939-1455},
  doi = {10.1037/0033-2909.119.2.254}
}

@article{pekrunControlValueTheoryAchievement2006,
  title = {The {{Control-Value Theory}} of {{Achievement Emotions}}: {{Assumptions}}, {{Corollaries}}, and {{Implications}} for {{Educational Research}} and {{Practice}}},
  shorttitle = {The {{Control-Value Theory}} of {{Achievement Emotions}}},
  author = {Pekrun, Reinhard},
  year = {2006},
  month = dec,
  journal = {Educational Psychology Review},
  volume = {18},
  number = {4},
  pages = {315--341},
  issn = {1573-336X},
  doi = {10.1007/s10648-006-9029-9},
  urldate = {2025-10-04},
  langid = {english}
}

@article{swellerCognitiveArchitectureInstructional2019,
  title = {Cognitive {{Architecture}} and {{Instructional Design}}: 20~{{Years Later}}},
  shorttitle = {Cognitive {{Architecture}} and {{Instructional Design}}},
  author = {Sweller, John and {van Merri{\"e}nboer}, Jeroen J. G. and Paas, Fred},
  year = {2019},
  month = jun,
  journal = {Educational Psychology Review},
  volume = {31},
  number = {2},
  pages = {261--292},
  issn = {1573-336X},
  doi = {10.1007/s10648-019-09465-5},
  urldate = {2025-10-04},
  langid = {english}
}

@incollection{winneStudyingSelfregulatedLearning1998,
  title = {Studying as Self-Regulated Learning},
  booktitle = {Metacognition in Educational Theory and Practice},
  author = {Winne, Philip H. and Hadwin, Allyson F.},
  year = {1998},
  series = {The Educational Psychology Series},
  pages = {277--304},
  publisher = {Lawrence Erlbaum Associates Publishers},
  address = {Mahwah, NJ, US},
  isbn = {978-0-8058-2481-0 978-0-8058-2482-7}
}

\end{document}